\documentclass{article}
\usepackage{graphicx}
\begin{document}
\title{Spinless particle in rapidly fluctuating random magnetic field}
\author{V.G.Benza and B.Cardinetti\\
Dipartimento di Fisica, Universita' di Milano\\
Via Celoria 16, 20133 Milano\\
also INFM, unita' di Milano}
\maketitle
\begin{abstract}
We study a two-dimensional spinless particle in a 
disordered gaussian magnetic field
with  short time fluctuations, by means of 
the evolution equation for the density matrix $<{\bf x}^{(1)} 
|\hat{\rho} (t)|{\bf x}^{(2)}>$; in this description the two 
coordinates are associated with the retarded and advanced 
paths respectively. 
The static part of the vector potential correlator is assumed 
to grow with distance with a power $h$; the case $h\,\,=\,\,0$
corresponds to a $\delta$-correlated magnetic field,
and $h\,\,=\,\,2$ to free massless field. 
The value $h\,=\,2$ separates two different regimes, 
diffusion and logarithmic growth respectively.
When $h\,<\,2$
 the baricentric coordinate ${\bf r} = 
(1/2)({\bf x}^{(1)} + {\bf x}^{(2)})$ diffuses with a 
coefficient $D_{r}$ proportional to $x^{-h}$, where 
${\bf x}$ is the relative coordinate: 
${\bf x}\,=\,{\bf x}^{(1)}-{\bf x}^{(2)}$. 
As $h\,>\,2$ the correlator of the magnetic field 
is a power of distance with positive exponent;
then  the coefficient $D_{r}$
scales as $x^{-2}$. 
 The density matrix is  a function of $r$ and ${x^{2} \over
t}$,and its width in $r$ grows for large times proportionally 
to  $log({t \over x^{2}})$.

\end{abstract}
\section{Introduction}
In recent years various studies have been devoted 
to  propagation in the presence of 
a stochastic drift term.
The most known example comes from fluid dynamics: 
the transport of a passive scalar in a 
velocity field. 
It has been shown that this problem can be solved 
in various physically relevant situations. 
 In particular Gawedzky and Kupiainen \cite{GK}
gave a solution in the three dimensional 
case by means of an explicit resummation of the 
perturbative series.
 Correlators of the 
scalar field of any order can be obtained.
The basic ingredient  is the 
assumption of zero correlation time for the velocity, 
which allows for an effective evolution generator.
Quotation  of previous  work on the passive 
scalar can be found, 
e.g., in Ref.\cite{FKLM}. 
A second, related example is 
 quantum motion in a disordered magnetic 
field. This topic has been raised in various 
physical contexts:   
Mott-Hubbard systems \cite{IoffeLarkin}, 
 vortex lines in superconductors 
\cite{Feigel}, \cite{C},
 Quantum Hall Effect at filling 
factor $1/2$  \cite{KZ},\cite{HLR}.
In  the last case  
the magnetic field is static with zero mean and 
numerically there is evidence of a delocalization 
transition \cite{AHK},\cite{KWAZ},\cite{AZ}, at odds 
with a supersymmetric 
treatment leading to a sigma model with unitary symmetry\cite{AMW}.
Fluctuations in 
the topological density have also been called
responsible for the delocalization transition 
\cite{ZA}. Recent work, in  examining the connection 
between passive scalar and quantum particle in a magnetic field,  
 related delocalized states with
random antisymmetric disorder,
but also  with symmetric 
disorder, provided  a peculiar sublattice 
decomposition is allowed
\cite{MIW}. Quantum mechanics 
with an imaginary magnetic field and a disordered potential, 
 introduced by
 Hatano and Nelson \cite{HN}, recently attracted much attention, 
(see, e.g., Efetov \cite{E}, showing how 
nonhermiticity can sustain extended states in spite of
disorder.
In a previous work on
static magnetic disorder we have shown that it translates, at 
 the semiclassical level, into an  
imaginary (nonlocal in time) magnetic field.
We further interpreted the effective action in terms of  Coulomb gas, 
and discussed how its strong coupling regime can enhance 
phase coherence among trajectories
  \cite{CB}.
Static disorder cannot be studied in terms of an effective 
time evolution generator, due to the  non-locality of the 
effective action.
Time-dependent disorder, in the limit of zero-correlation time, 
can instead be studied: one can think of a low energy particle 
moving in a sea of highly energetic ones, each carrying a magnetic 
flux line, or to a particle, as, e.g., a vortex, in a fastly 
varying effective magnetic field.
Cooperon dynamics with time-dependent magnetic fluctuations  
has been studied 
 by Aronov and Wolfle \cite{AW}, in the context of high-$T_c$ 
superconductors. 
In  the present work we  analize
 rapidly fluctuating magnetic disorder on an 
otherwise free particle: our aim is to determine 
whether the ballistic behavior is frustrated in such a case, 
and what kind of presumably slower propagation sets in.
By taking the correlation length of the fluctuations as 
the large scale of the system  
we  establish the effective (annealed) dynamics for 
various power-law disorder correlators.
In the second Section we determine the time evolution generator 
for the density matrix
from the Feynman path integral.
This is  a two-particle description, since 
both retarded and advanced paths must be taken into account.
Averaging over disorder leads to self-interaction as well as to 
mutual interaction between the paths: the former amounts to 
a renormalization of the single particle dynamics 
(i.e. it  influences  the quantum amplitude), 
while the latter, referring to interference between counterpropagating 
paths, is essential to quantum probability. 
In the third Section we discuss 
the particle-antiparticle relative dynamics,  
which is related with the quantum fluctuations 
in the momentum space.
In Section 4 we analize the full problem and give the main results 
on the particle diffusion and subdiffusion.
In Section 5 we make some comparison with 
similar behaviors in different physical contexts.
 
\section{Density matrix evolution}
 We assume that the correlation time 
of magnetic fluctuations can be neglected and take  
 the following gaussian 
correlator for the vector potential, 
 in the transverse gauge:  
\begin{equation}
 D_{\alpha,\beta} ( {\bf k} ) \cdot \delta (t) \equiv 
< A_{\alpha} A_{\beta} > ( {\bf k} \, ,t) =  \Delta  
\delta^{T}_{\alpha, \beta} ( {\bf k} ) 
{ 1 \over ( k^{2} + k^{2}_{0})^{1 + h/2} } \delta (t),
\label{corr}
\end{equation}
where $ k_{0}^{-1} $ is the correlation length,  
$ h $ is positive and $ \delta^{T}_{\alpha,\beta} 
\equiv \delta_{\alpha,\beta} - (k_{\alpha} \,\, k_{\beta})/k^{2}$.
We will take into account the range $ 0\,\,<\,\,h\,\,<\,\,4$;
\,\, the limit $ h \rightarrow 0 \,\, , k_{0} \rightarrow 0$ 
gives a magnetic field $B$ delta-correlated 
in space,
\,\,  $ h = 3$ can be associated with 
the anomalous skin effect \cite{AW};\,\,
 $h\,\,=\,\,2$ gives a free massless $B$ and,  
apart from logarithmic corrections,
corresponds to the enstrophy cascade in two-dimensional 
turbulence (with the vector potential playing the role of 
velocity).
 Notice that as  $h\,>\,2$ the correlator of $B$ is 
a power of distance with positive exponent.
We write the Feynman 
path integral representation  
for the product of amplitudes:
\begin{equation}
F({\bf x},{\bf y};{\bf x}',{\bf y}';t) \equiv
< {\bf x} \mid U^{(1)}(t) \mid {\bf y} > < {\bf x}' \mid 
U^{(2)}(t) \mid {\bf y}' > ^{ \ast},
\label{prod.ampl}
\end{equation}
where $U^{(i)}(t) \,, i =1,2$ are the evolution operators 
of two identical copies of single particle systems in the presence 
of the magnetic field.
The time-evolved density matrix is  given by:
\begin{equation}
<{\bf x}|\hat{\rho} (t)|{\bf x}'>\,\,=\,\,
\int d{\bf y} d{\bf y}'F({\bf x},{\bf y};{\bf x}',{\bf y}';t) 
<{\bf y}|\hat{\rho}
(0) 
|{\bf y}'>
\label{ev}
\end{equation}
 The representation of Eq. \ref{prod.ampl}
involves a two-particle action: advanced and retarded path 
respectively. 
 Averaging over the vector potential couples the paths and generates 
 the following 
 effective 
lagrangian $ \mathcal{L}_{eff}
( {\bf X} \,,\dot{ {\bf X} })\,, {\bf X} \equiv ( {\bf x}^{(1)} 
\,, {\bf x}^{(2)} ) $ :
\begin{equation}
\mathcal{L}_{eff} = {m \over 2} \dot{ \bf X} \sigma_{3}
\dot{ \bf X} + {i \cdot g^{2} \over 2 \cdot \hbar} 
\dot{ \bf X} \sigma_{3} \hat{D} \sigma_{3} \dot{ \bf X} \equiv
{1 \over 2} \dot{ \bf X} \hat{M} \dot{ \bf X}.
\label{Leff}
\end{equation}
Here $ g = {e \over c} $, the Pauli matrix 
$ \sigma_{3}$ acts on the 
particle indices and the (4X4) matrix $ \hat{D} \equiv 
\hat{ D} ^{(i,j)}\,\,(i,j\,\,=\,\,1,2) $ corresponds to 
the static part of the 
correlator defined in Eq. ~\ref{corr}: $ \hat{D} ^{(1,1)} 
= \hat{D}^{(2,2)} = D_{\alpha, \beta} ( {\bf x}^{(1)} = {\bf x}^{(2)})\, 
; \hat{D} ^{(1,2)} = \hat{D} ^{(2,1)} = 
D_{\alpha, \beta} ( {\bf x}^{(1)} - {\bf x}^{(2)} )$;
$( \alpha,\, \beta \equiv x,\,y)$.
The evolution of the density matrix is given by Eq. \ref{ev}, where 
 $F$ is converted into its average over disorder; in the effective action
 the imaginary part of $\hat{M}$ is obviously related with dissipation.
Since, contrary to the static case (see, e.g.,Ref. \cite{CB}), we have 
locality in time, it is possible to extract
  a time evolution generator $\hat{H}$ from Eq. \ref{ev}. 
This can be performed with no ambiguity in a flat metrics, 
while in general operator ordering 
prescriptions are needed.
As discussed at length in Ref. \cite{AW},
if one evaluates the metrics 
at the midpoint between the  initial 
and final configurations, and consistently normalizes 
the intermediate gaussian integral,
one obtains a symmetrized 
generator.
We have:
$ \hat{H} = {1 \over 2} ( \nabla \hat{ F} ^{-1} \nabla )_{s}$ 
where $ \hat{F} = { 1 \over i \hbar } \hat{ M }$  
and $ ( \nabla \hat{A} \nabla )_{s} \equiv {1 \over 4} 
( \nabla \nabla \hat{A} + 2 \nabla \hat{A} \nabla +
\hat{A} \nabla \nabla )$. Here the sum over indices is understood 
 and the gradient is with respect to ${\bf X}$.
The kernel $\hat{ F} ^{-1}$ is given by:
\begin{eqnarray}
\hat{ F} ^{-1} & =& ( {i \hbar \over m } \sigma_{3} 
+ ( {g \over m} )^{2} \hat{D} ) \cdot \hat{ G}^{-1}  \\
 G_{\alpha, \beta} & = & \delta_{\alpha, \beta} 
+ ( { g^{2} \over m \cdot \hbar } )^{2} 
\cdot \big[  D^{2}_{\alpha, \beta}( {\bf 0} ) - 
 D^{2}_{\alpha, \beta}( {\bf x}^{(1)} - {\bf x}^{(2)})\big]
\nonumber 
\label{invF}
\end{eqnarray}
This simple form results from $D_{\alpha, \beta}$ being diagonal 
when its argument is equal to $ {\bf 0}$.
The operator $\hat{G}$ is diagonal in the particle indices and, 
as one easily verifies, is even under particle exchange; it obviously 
commutes with $\hat{D} \equiv \hat{D} ( {\bf x}^{(1)} - {\bf x}^{(2)})$.
By taking into account the transversality condition  
 one ends up with the following evolution 
equation for the density matrix $\rho ( {\bf x}^{(1)}\,,
 {\bf x}^{(2)}\,; t)$ in coordinate representation:
\begin{eqnarray}
\label{H}
 {\partial \rho \over \partial t} & =& \hat{H} \rho; \nonumber  \\
 \hat{H} & =& {i \hbar \over 2 m} (G^{-1})_{\alpha, \beta} 
 (\partial ^{(1)}_{\alpha} \partial ^{(1)} _{\beta}  
-  \partial ^{(2)}_{\alpha} \partial ^{(2)} _{\beta}) \nonumber \\ 
&&+ {1 \over 2} ({g \over m})^{2} D^{0}_{\alpha, \gamma} 
(G^{-1})_{\gamma , \beta} (\partial ^{(1)}_{\alpha} 
\partial ^{(1)}_{\beta} + \partial ^{(2)}_{\beta} 
\partial ^{(2)}_{\alpha}) \nonumber \\     
&&+ {1 \over 2} ({ g \over m})^{2} D^{x}_{\alpha, \gamma}
(G^{-1})_{\gamma , \beta} (\partial ^{(1)}_{\alpha}  
\partial ^{(2)}_{\beta} + \partial ^{(2)}_{\alpha} 
\partial ^{(1)}_{\beta}) \nonumber \\
&&+ {i \hbar \over 2 m} 
\big [ \partial _{\alpha} ( G^{-1})_{\alpha, \beta} \big ]
(\partial ^{(1)}_{\beta} + \partial ^{(2)} _{\beta}) \\
&&+ {1 \over 2} ({ g \over m})^{2} (D^{0} - D^{x})_{\alpha, \gamma}
\big[\partial_{\alpha}(G^{-1})_{\gamma, \beta}\big](\partial ^{(1)}_{\beta} 
 - \partial^{(2)}_{\beta}) \nonumber \\
&&+ {1 \over 4} ({ g \over m })^{2} \partial _{\gamma}
\big[( D^{0} - D^{x})_{\alpha, \beta} \partial _{\alpha} 
(G^{-1})_{\beta, \gamma}\big] \nonumber     
\end{eqnarray}
Here the partial derivatives, unless labeled with the particle 
index, are taken with respect to the relative coordinate 
$ {\bf x} \equiv {\bf x}^{(1)} - {\bf x}^{(2)}$ and $D^{0} \,, D^{x}$ 
refer to the static part of the correlator at separation $ {\bf 0} $ 
and $ {\bf x} $  respectively.
Notice that  the particle mass is
 renormalized, and
an effective magnetic field (proportional to 
$({g \over m})^{2}$) acts with opposite signs
on the two trajectories.
 The remaining terms are dissipative, 
of diffusion, potential and drift type:  obviously this acts as an
imaginary magnetic field (proportional 
to ${\hbar \over m}$) acting identically on both trajectories. 
If, in the original hamiltonian, one disregards the ${\bf A}^{2}$
term, one can average the two-particle evolution operator
emerging from Eq. \ref{prod.ampl}. 
 Following the lines of Ref. \cite{GK}, one starts with the Neumann 
series for this operator,taking as  perturbation  the 
${\bf A} \cdot {\bf p}$ coupling, then uses Wick's theorem and 
averages each term. 
The  resulting series can be reexponentiated 
to an effective two-particle generator, which coincides with 
Eq. \ref{H}, but with
$\hat{G}\,\,=\,\,\hat{1}$.
This case is fortunate since the transversality condition guarantees 
that the 
generator $\hat{H}$ does not depend on ordering prescriptions.

\section{Particle-antiparticle relative dynamics} 
Upon isolating the dynamics of
  the relative coordinate, as  from the center of mass 
of the system, we  study the  
 interaction between the retarded and advanced 
paths. 
If only the relative
coordinate survives, the operator $\hat{H}$ reduces to: 
\begin{eqnarray}
\label{Hrid}
\hat {H} &=& ({ g \over m})^{2} (D^{0} - D^{x})_{\alpha ,\gamma} 
(G^{-1})_{\gamma , \beta} \partial_{\alpha} \partial_{ \beta} \nonumber\\
&& + ({ g \over m})^{2}(D^{0} - D^{x})_{\alpha, \gamma} 
\big[\partial_{\alpha}(G^{-1})_{\gamma , \beta}\big] \partial_{\beta}\\
&& + {1 \over 4} ({ g \over m})^{2} \partial _{\gamma}
\big[(D^{0} - D^{x})_{\alpha, \beta} \partial_{\alpha} 
(G^{-1})_{\beta , \gamma}\big] \nonumber
\end{eqnarray}
 
We now show that Eq.  \ref{Hrid} in the case of long range fluctuations 
$(k_{0}\,\,<<\,\,1)$ further simplifies, so that  the drift and 
curvature contributions can be neglected.
In order to proceed, it is first necessary to examine the disorder
correlator in that regime.
In space coordinates the static part has the form:
\begin{eqnarray}
\label{Bessel}
D_{\alpha, \beta}({\bf x})& = & ({ \Delta \over 4 \pi})(\mathcal{I} _{0}(x)
\cdot \delta _{\alpha, \beta} 
+ \mathcal{I} _{2} (x) \cdot \Sigma _{\alpha , \beta}) \\
\Sigma_{1,1} & = & - \Sigma_{2,2} = \cos ( 2 \psi )\,;
\Sigma_{1,2} = \Sigma_{2,1} = \sin ( 2 \psi ) \nonumber \\
\mathcal{I}_{i} ( x) & = & \int ^{\infty} _{0}\!
{k\,dk \over ( k^{2} + k^{2}_{0})^{1 + h/2}}\, J_{i}(k\,x) \nonumber
\end{eqnarray}   
where $x$ and $\psi$ are the polar coordinates of ${\bf x}$ and 
the $J_{i}$s are Bessel functions. The small $k_{0}$ behavior of
Eq. 
\ref{Bessel} is readily obtained:
\begin{eqnarray}
\label{RT}
D_{\alpha, \beta} & = & ({\Delta \over k^{h}_{0}})
({1 \over 4\, \pi \,h} \delta_{\alpha, \beta} + 
\mathcal{A}_{\alpha, \beta}) \\
\hat{ \mathcal{A}} & = & ({ \xi \over 2})^{h} \hat{R}\,\,\, , 0 <\,h\,<2
\,\,\,\,\, \xi \equiv k_{0} \cdot x \nonumber \\
\hat{ \mathcal{A}} & = & ( \xi )^{2} \hat{T}\,\,\,\,, 2<\,h\,<4 \nonumber \\
\hat{R} & = & {1 \over 8\, \pi} { \Gamma (1- h/2) \over \Gamma (1 +
h/2)}\big[ - ( 1 + 2/h) \hat{1} + \hat { \Sigma}\big] \nonumber \\
\hat{T} & = & { 1 \over 8 \, \pi} {1 \over h \cdot (h -2)}
( - \hat{1} + { 1 \over 2} \hat{ \Sigma}) \nonumber
\end{eqnarray}
The explicit form of $ \hat{G} $  is then;
\begin{equation}
\label{G}
\hat{G} = \hat{1} - ({g^{2} \over m \cdot \hbar} { \Delta \over
k^{h}_{0}})^{2}
 \hat{\mathcal{A}} ({1 \over 2 \pi \hbar} \hat{1} + \hat{\mathcal{A}})
\end{equation}
The subtracted correlator has a power-law behavior, and is $O(k_{0}^{0})$ 
 in the limit $k_{0} \to 0$ when $h\,\,<\,\,2$:
\begin{equation}
\label{Dsottr}
\hat{D} ^{0} - \hat{D} ^{x} =  - { \Delta \over k_{0}^{h} }  
\hat{\mathcal{A}}
\end{equation}
The ratio $ { \Delta \over k_{0}^{h}} $ has the dimensions of a
momentum, let us call it $ \Pi $. 
The order of magnitude 
of the  $ \bf{A} ^{2} $ contribution can be estimated 
 by means of the adimensional 
coupling constant $ {{ r_{e} \cdot \Pi} \over \hbar } $, where 
$ r_{e} $ is the classical radius of the electron $ ( r_{e} 
\equiv { g^{2} \over m}) $.
We define the large fluctuations regime, where 
in $\hat{G}$ we can disregard the identity with respect to the second 
term:
\begin{equation}
\label{largefluc}
{ g^{2} \cdot \Delta \over m \cdot \hbar \cdot k_{0}^{h}}\,\, 
\equiv { r_{e} \cdot \Pi \over \hbar}\,\, >> 
\,\, 1
\end{equation} 
Notice that this relates the amplitude of the fluctuations 
(associated with $\Delta$) with  the correlation length 
 $k^{-1}_{0}$. The kernel$\hat{F}^{-1}$, a function of $\Pi$, 
$h$ and
of the quantum flux unit ${\hbar \over g} \equiv { \Phi_{0} 
\over (2 \pi)}$, reduces to:
\begin{eqnarray}
\label{invmass}
\hat{F} ^{-1} & = & ({ \hbar \over g})^{2}{ k_{0}^{h} \over \Delta \gamma} 
\cdot ( \hat{1} - { 1 \over \gamma} \hat{\mathcal{A}}) \\
\gamma & \equiv & {1 \over 2 \cdot \pi \cdot h} \nonumber
\end{eqnarray}
It is easily verified that in both cases $ ( h \, <2\,\, , h \, > \,
2) $
 the matrix $ \hat{\mathcal{A}} $ satisfies the 
transversality condition; this implies that the generator is, 
independently of ordering 
 prescriptions:
\begin{equation}
\label{Hw}
\hat{H} = {1 \over 2} ({ \hbar \over g})^{2} { 1  \over \Pi 
\cdot \gamma}( \delta_{\alpha, \beta} - { 1 \over \gamma} \mathcal{A}
_{\alpha, \beta}) \partial _{\alpha} \, \partial_{\beta}
\end{equation} 
 The dominant contribution is pure  diffusion with a 
coefficient $O(k_{0}^{h})$, while the  first 
correction has scaling form,
and in both cases it corresponds to
 a faster spreading.  When 
$ 0\,\,<\,\, h\,\, < \,\, 2 $ , the correction
 is of order $ k_{0}^{2h} $ and, taken alone, would give 
superdiffusion $ ( t \approx x^{2 - h}) $;   when $ 2\,\,<\,\,h
\,\,< \,\,4 $, it 
has order $ k_{0}^{2+h} $ with  $ t \approx (\ln x)^{2} $.
A complete analysis of Eq. \ref{Hw} is needed in order to  
understand the interplay between diffusion and this faster 
mechanism; we leave it to future work.
We finally  discuss the regime 
 of weak fluctuations $ ({ r_{e} \cdot
\Pi \over \hbar} \,\,<<\,\,1) $. When in $\hat{G}$ we disregard 
the second term, 
 $\hat{F} ^{-1}$ has 
a  scaling form:
\begin{equation}
\label{Hww}
\hat{F} ^{-1} = - ({g \over m})^{2} \Pi \hat{ \mathcal{A}} \\ 
\end{equation}
Recall that while in the $h\,\,<\,\,2$ range this is $O(k^{0}_{0})$, in the
complementary range it diverges as $k^{2 -h}_{0}$ in the limit 
$k_{0} \to 0$.
Corrections in the small fluctuation parameter come 
 from the expansion of $\hat{G} ^{-1}$.
 Since only the dominant term preserves the transversality
 condition,  the operator $\hat{H}$  has 
a drift  as well as a curvature term. The dominant part 
 gives rise to superdiffusion $( x \approx t ^{1/(2 -h)} )$ 
when $0 \,\, < \,\,h \,\, <\,\, 2$ and time exponential behavior (as 
 previously illustrated) when 
$ 2 \,\, < \,\,h\,\, < \,\, 4 $. This amounts to  
disregarding the $\bf A^{2}$ term .
The  generator corresponds then to the one 
obtained for the second order correlator of 
a passive scalar  
 by Gawedzki and Kupiainen  \cite{GK}. 
We stress in conclusion that the scaling form of the 
disorder
does not transfer to the generator, due to the fluctuations 
of  ${\bf A}^{2}$; this reduces the 
strong tendency of the 
two trajectories towards separation, from 
fluctuation-supported superdiffusion 
to diffusion.
 
\section{General case}
Let us write  equation \ref{H} in terms of the relative
coordinate and of the baricentric coordinate $ { \bf r} \equiv 
 {1 \over 2} ( { \bf x}^{(1)} + { \bf x}^{(2)} ) $ :
\begin{eqnarray}
\label{H1}
{ \partial \rho \over \partial t} & =& \hat {H} \rho \\
\hat {H} & = & { i \hbar \over 2m} (G^{-1})_{\alpha, \beta}
({\partial \over \partial r_{\alpha}} {\partial \over \partial x_{\beta}} + 
{\partial \over \partial x_{\alpha}} {\partial \over \partial r_{\beta}}) + 
{ i \hbar \over 2m} 
\big[ {\partial \over \partial x_{\alpha}} ( G^{-1})_{\alpha,
\beta}\big]
 {\partial \over \partial r_{\beta}} \nonumber\\
&& + {1 \over 4} ({ g \over m})^{2}
\big[(D^{0} + D^{x})G^{-1}\big] _{\alpha,\beta} 
{\partial \over  \partial r_{\alpha}} {\partial \over \partial r_{\beta}} + 
({g \over m})^{2}
\big[( D^{(0)} - D^{(x)})G^{-1}\big]_{\alpha, \beta} 
{\partial \over \partial x_{\alpha}} {\partial \over \partial
x_{\beta}} 
\nonumber\\
&& + ({g \over m})^{2}
\big[{\partial \over \partial x_{\alpha}}( D^{0} - D^{x})G^{-1}
\big]_{\alpha, \beta} {\partial \over \partial x_{\beta}} + 
{1 \over 4} ({ g \over m})^{2} {\partial \over \partial x_{\alpha}} {\partial 
\over x_{\beta}}\big[(D^{0} - D^{x})G^{-1}\big]_{\alpha, \beta} \nonumber 
\end{eqnarray}
 Here 
 the limit $ k_{0} \rightarrow 0$ gives a finite nonzero result provided 
that $h \,\,<\,\,2$:  as discussed in the previous section, the 
 diffusion term for ${\bf x}$
is  $O( k_{0}^{h})$, similarly the other terms go to zero in the limit.
From  Eq.  \ref{H1} we get then:
\begin{equation}
\label{Hsurv}
\hat{H} \rightarrow -{1 \over 4 \Delta}({\hbar \over g})^{2}({x \over
2})^{-h} (\hat{R} ^{-1})_{\alpha,\beta} 
{\partial \over \partial r_{\alpha}}
{\partial \over \partial r_{\beta}}
\end{equation}
If one  disregards the anisotropy, one easily concludes that this 
implies an evolution through the combination 
${{r^{2} \cdot x^{h}} \over t}$. 
 Let us consider:
\begin{equation}
\label{Happrox}
\hat{H} \approx \mathcal{D} (x/2)^{-h} \nabla^{2}_{{\bf r}}\nonumber
\end{equation}
where $\mathcal{D}$  is constant.  
Let us take as initial condition a
gaussian in both particle coordinates with equal width:
\begin{equation}
\label{roin}
\rho ({\bf r} , {\bf x}) = ({w \over 2 \pi})^{2} exp(-w(r^{2} +
(x/2)^{2}))
\end{equation} 
The solution is:
\begin{equation}
\label{rout}
\rho( {\bf r} ,{\bf x},t) = {w \over 2 (2 \pi)^{2}} 
{1\over {1/ 2w + 2 \mathcal{D} (2/x)^h t}} 
exp\big[-(w/4)x^{2} - (1/2){ r^{2} \over {(1/2w) + 2 \mathcal{D}
(2/x)^{h} t}}\big]
\end{equation}

 The Wigner function, which in the 
classical limit goes into the phase space distribution, can be
evaluated at zero 
momentum; it
 is defined as:
\begin{equation}
\label{Wig}
W({\bf r},{\bf k},t) \equiv {1 \over (2 \pi)} \int d^{2} {\bf x} 
exp(-i {\bf k}\cdot {\bf x})\rho ( {\bf r}, {\bf x},t) \nonumber
\end{equation}
 In the large time limit we disregard $1/(2w)$ with respect to
$2 \mathcal{D}(2/x)^{h}t$ and obtain:
\begin{eqnarray}
\label{sol}
W({\bf r},{\bf k}= {\bf 0},t)&=&{1 \over (2 \pi)^{2}}
{w \over    { 2^{h +2} \mathcal{D} t}}\int ^{\infty} _{0} dx \cdot x^{h+1}
 exp(-(w/4) x^{2} - Q x^{h}) \nonumber\\
& \approx & 
  {w \over (2 \pi)^{2}} { 1 \over  { 2^{h+2} \mathcal{D} t}}
\int^{1/(w)^{1/2}}_{0} dx x^{h+1} exp(-Q x^{h}) \nonumber \\
Q  \equiv   r^{2} \cdot (2^{h+2} \mathcal{D} t)^{-1} \,\,\,&,&
\mathcal{D} t \,\,>\,\, { 1 \over {(2 w^{1/2})^{2+h}}}\,\,;\,\,
{ r^{2} \over { \mathcal{D} t}}\,<\,2^{h} \cdot w^{h/2}
\nonumber 
\end{eqnarray}
The integral can be explicitly written in terms of the truncated gamma 
function, and turns out to be a function of 
${r^{2} \over {\mathcal{D} t}}$.
The ${\bf k}$ dependence of $ W$  will not be discussed 
 in detail here.
Notice only  that, for real $p$,  $exp(-\,a \cdot|p|^{h})$ is the Fourier 
transform of the Levy distribution
$L_{h}(z) \approx {1 \over {z^{(h+1)}}}\,\,(z\,\,>>\,\,1)$   
\cite{MOW}. 
The density matrix 
displays this stretched exponential for every initial 
condition provided it is gaussian-shaped 
 in ${\bf r}$.
It is reasonable to assume that the large ${\bf k}$ behavior of $W$
 is 
 only marginally influenced by the initial ${\bf x}$ 
dependence\,\,$(x\,<<\,1)$,
if  sufficiently smooth.
The Wigner function  reduces then to 
 the $Q$-derivative of 
the Levy distribution in momentum space , where the derivative absorbs 
the factor $x^{h}$, from the normalization in 
${\bf r}$ .
 Thus the relative motion can  be depicted as a sequence of jumps 
and stopovers in momenta, giving rise to a hyerarchy of clusters. 
So far we have been dealing with
 the  $k_{0} \to 0$ limit; 
as $k_{0}$ is small but different from zero we must take into account
other terms, starting with 
 the first correction, which 
 gives pure diffusion in ${\bf x}$ (see the 
previous section). 
We  discuss a situation of this sort in the regime
 $2\,\,<\,\,h\,\,<\,\,4$. Now $D_{r}$
 is $O(k_{0}^{h-2})$; this, together with the next to 
leading term gives  
 ( $ {{k_{0}^{h}} \over {(\xi)^{2}}}\,\,<<\,\,1$): 

\begin{equation}
\label{2<h}
\hat {H} \approx ({\hbar \over g})^{2} {1 \over \Pi} \cdot \big[
 (-1/4) {1 \over {(k_{0}x)^{2}}} (\hat{T}^{-1})_{\alpha, \beta}
{\partial \over \partial r_{\alpha}}{\partial \over \partial r_{\beta}} 
+ (2 \pi h) \nabla^{2}_{\bf x} \big]
\end{equation}
 Disregarding anisotropy
and Fourier transforming with respect to ${\bf r}$ we
get a particle in a centrifugal barrier:

\begin{eqnarray}
\label{centr} 
{\partial \rho \over \partial \tau}&=&
\big[- ({q \over \xi })^{2} + \nabla^{2}_{{\bf
\xi }}\big] \rho\\
\tau&=&({\hbar \over g})^{2}{{k_{0}^{h} 2 \pi h}\over {
\Delta }} \cdot t;\nonumber\\
q^{2}&=&(4/3)(h-2){K^{2} \over k_{0}^{2}},\nonumber
\end{eqnarray}
where ${\bf K}$ is the momentum with respect to ${\bf r}$.
 Let us
 search for a similarity solution, 
$\rho = f(u,q)\,\,,u \equiv {\xi^{2} \over \tau}$, as suggested 
by the scaling form of the equation; the result is:
\begin{equation}
\label{hypconf}
 f'' + ({1 \over u} + 1/4)f' - ({q \over 2u})f =0
\end{equation}
 Two independent solutions (when $q$ is not an integer) are:
$f(u,q) = u^{s}F(s,2s+1;-u/4)$, where $s\,\,=\,\,q/2\,\,,-q/2$
and F is the confluent hypergeometric function.
In order to avoid singular behavior at small $u$ 
we exclude the negative $s$ case.
We numerically computed the  
 Fourier transform , in order to
 reconstruct the ${\bf r}$ dependence; 
 in Fig.1 we show $\rho(r\,,\,u)$ for a set of values $u \,<\,1$, i.e. 
in the large time regime. We find that the width in $r$  
 increases, as $u$\,\,approaches zero,  linearly in the 
logarithm of $u^{-1}$, up to four orders of magnitude (see Fig.2).      
  Disregarding  the  next to leading term  
would result, similarly to the previous case,  
in the combination ${{r^{2}\cdot x^{2}} \over t}$.
 
\section{Conclusions}
We studied the two-dimensional dynamics of a  
particle  in the presence of a stochastic magnetic
field in the fast fluctuation regime. We have found 
a transition
 by varying the exponent of the disorder correlator.
 Disregarding the baricentric 
motion, we find that diffusion is the dominant 
behavior in the $ k_{0} \,\, << \,\, 1$ regime. Superdiffusion arises in  
 the first correction, with the scaling $t \approx 
x^{2-h}$ when $h\,\,<\,\,2$ and $ t \approx (\ln x)^{2}$ when 
$2\,\,<\,\,h\,\,<4$.  
The dominance of diffusion originates 
in the ${\bf A^{2}}$ term; if this term is neglected 
we obtain fluctuation-sustained superdiffusion with the previously 
mentioned 
power laws. In other words the tendency of the retarded and 
advanced trajectories to spread apart very rapidly (favored by 
 the ${\bf A} \cdot {\bf p}$ term) is partly stabilized by the
quadratic term.
In the general case, 
the limit $k_{0} \to 0$ can be performed exactly  
in the range $h\,\,<\,\,2$.  
The  solution  
(see Eq. \ref{rout}) shows diffusion in ${\bf r}$, with the 
dependence $r \approx (t/x^{h})^{1/2}$. 
 The Wigner function $W$ at zero momentum, i.e. 
 the average  over particle separation, 
describes 
diffusion 
($t \approx r^{2}$);
 at ${\bf k}$ large it appears in the form 
$W\, \approx  \,W({r^{2} \over t},{\bf
k})$, its ${\bf k}$ behavior being associated  
with  isotropic Levy flights in the momentum space.  
 Random walk in a quantum mechanical system was  
found in the Harper model at its critical point \cite{HA},\cite{GKP}.
It was also derived
from a two-dimensional anisotropic random lattice model,
describing edge propagation along the surface of a quantum Hall multilayer 
\cite{CD},\cite{BF},\cite{BFZ},\cite{MA}. In that case  backscattering 
is allowed  
in the direction of the magnetic field and neglected in the orthogonal
direction; when the latter is interpreted as time one has 
a one-dimensional chain with time-dependent hopping 
disorder where the quantum particle undergoes diffusion.
As long as hopping mimicks the presence of a magnetic field, 
this model can be taken as a one-dimensional version of ours.
Propagation  is further  inhibited when $2\,\,<\,\,h$;  
by including the leading and next to leading terms
we determined the Fourier transform with respect to 
${\bf r}$ of the density matrix 
(see Eq. \ref{hypconf} and arguments following it). From the  
large time
behavior of $\rho({\bf r},{\bf x},t)$ one extracts  
  $ r\, \approx \,log({t  \over x^{2}})$: this is a much 
slower spread 
than expected from
the leading term alone,  giving 
 $r \approx (t/x^{2})^{1/2} $.
This  ultraslow, logarithmic diffusion occurs in biased 
random motion on percolation clusters or globally 
isotropic fractals (see  \cite{BG} and references therein).
Quantum mechanically, it was  found 
by suitably perturbing 
at a single point an Anderson hamiltonian in the localized 
phase \cite{RJLS}.
When disorder is static, 
 \cite{MU},\cite{CL}, smoothly varying intense magnetic fields
 confine nearly free  states within a narrow one-dimensional 
region along the zero field lines, such lines thus providing a quantum 
percolation  network.
The present case 
can be more properly depicted in terms of
 abrupt jumps in the space of momenta,
with  loss of phase coherence.
This apparently results in 
 random walk when the correlator of the magnetic 
field $B$ decays with distance $(h\,\,<\,\,2)$ and in 
 logarithmic behavior  when it  grows 
$(2\,\,<\,\,h)$.

\section*{Acknowledgements}
We thank R.Artuso for pointing to our attention Refs. 
\cite{GKP} and \cite{RJLS}.
  
\section*{Figure captions}
Fig.1: Plot of the solution $\rho(r,u)$ (see Eq.\ref{hypconf}) 
at different values
of $u$ as $u$ approaches zero (large time behavior).\\
Fig.2: Width in $r$ of the solution versus $log(u)$.

\includegraphics{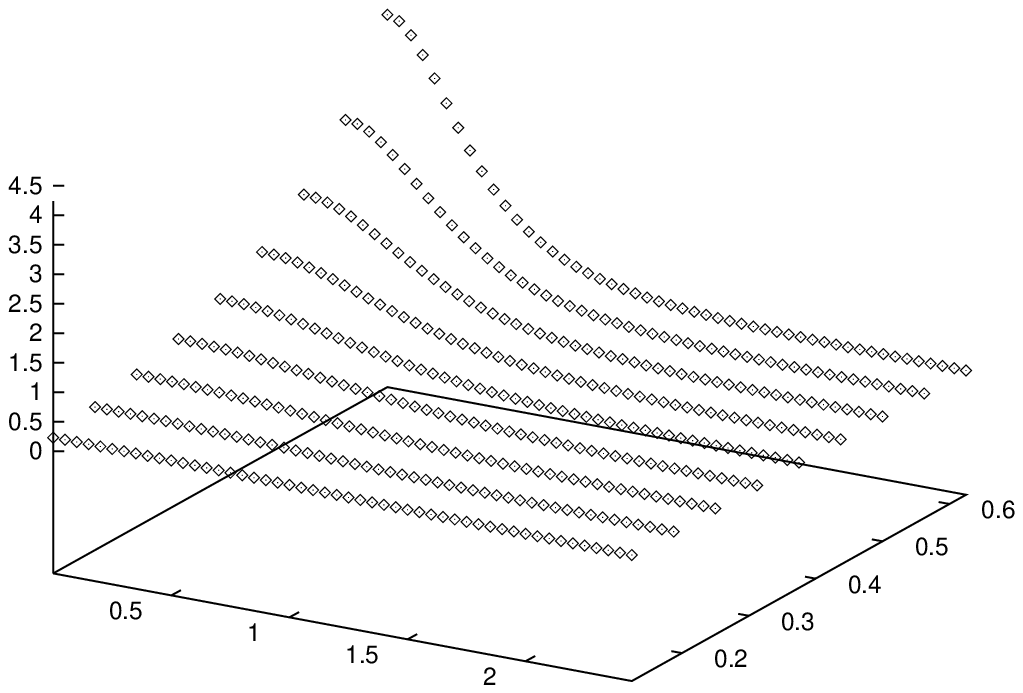}
\includegraphics{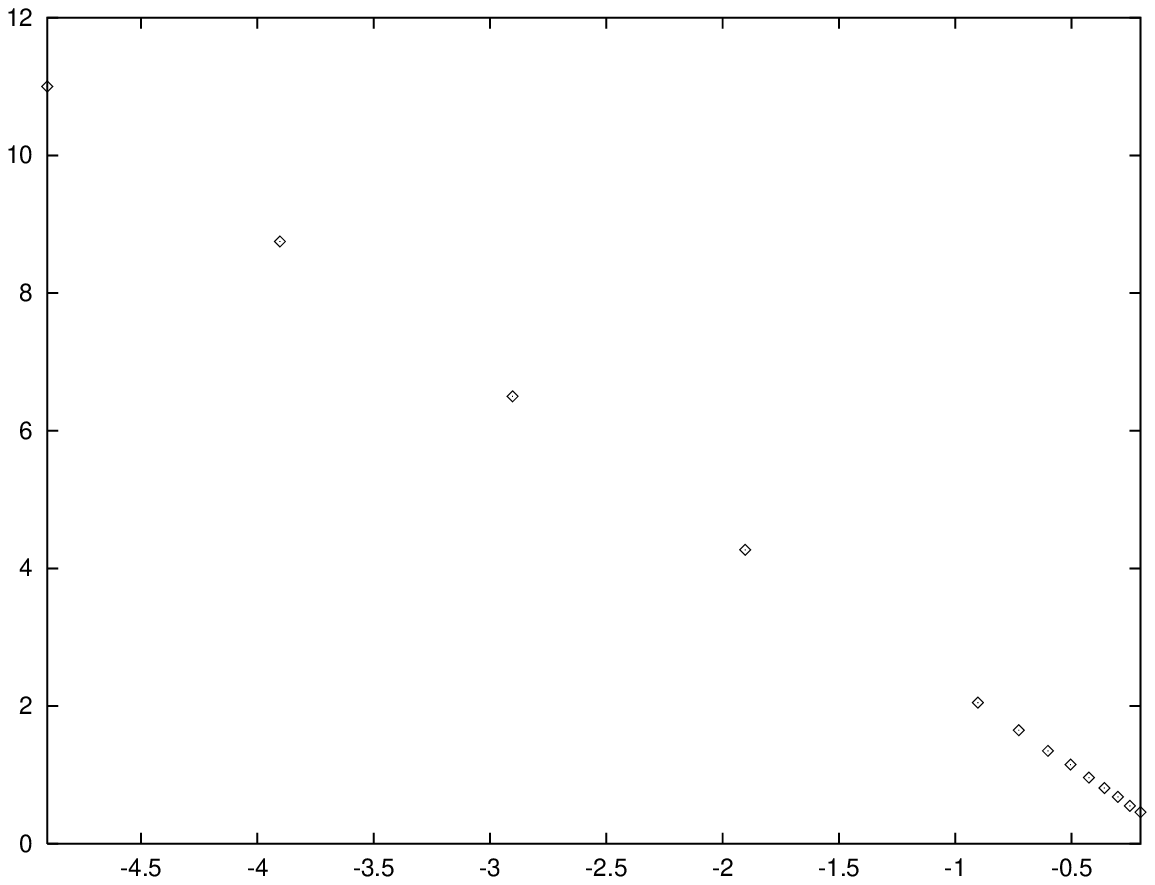}

\begin{thebibliography}{99}
\bibitem{GK} K.Gawedzki,A,Kupiainen: Universality in turbulence: an 
exactly soluble model. Lecture Notes of the $34^{th}$ School of 
Theoretical Physics in Schladming,Austria, March 4-11,1995.
\bibitem{FKLM}G.Falkovich,I.Kolokolov,V.Lebedev,A.A.Migdal,Phys.Rev.E{\bf
54},\\
4896(1996).
\bibitem{IoffeLarkin} L.B.Ioffe,A.I.Larkin, Phys.Rev.B {\bf 39},8988(1989).\\
 L.B.Ioffe,.V.Kalmeyer, Phys.Rev.B {\bf 44},750(1991).\\
 N.Nagaosa,P.A.Lee,Phys.Rev.Lett. {\bf 64},2450(1990).
\bibitem{Feigel}M.V.Feigelman,Physica A {\bf 168}, 319(1990).
\bibitem{C}E.M.Chudnovsky, Phys.Rev. {\bf 51},15351 (1995).
\bibitem{KZ}V.Kalmeyer,S.C.Zhang, Phys.Rev.B {\bf 46},9889(1992).
\bibitem{HLR} B.I.Halperin,P.A.Lee,N.Reed,Phys.Rev. b {\bf 47},7312(1993).
\bibitem{AHK}Y.Avishai,Y.Hatsugai,M.Kohmoto,Phys.Rev. b{\bf
47},9561(1993).
\bibitem{KWAZ}V.Kalmeyer,D.Wei,D.P.Arovas,S.C.Zhang,Phys.Rev. B {\bf 48},
11095(1993).
\bibitem{AZ}D.Z.Liu,X.C.Xie,S.Das Sarma,S.C.Zhang, Phys.Rev. B{\bf 52},
5858(1995).
\bibitem{AMW}A.G.Aronov,A.D.Mirlin,P.Wolfle, Phys.Rev.B {\bf
49},16609(1994).
\bibitem{ZA}S.C.Zhang,D.Arovas, Phys.Rev.Lett.{\bf 72},1886(1994).
\bibitem{MIW}J.Miller,J.Wang, Phys.Rev.Lett.{\bf 76},1461(1996).
\bibitem{HN}N.Hatano,D.Nelson, Phys.Rev.Lett.{\bf 77},570(1996).
\bibitem{E}K.B.Efetov,Cond-Mat/9706055, June 6th,1997.
\bibitem{CB}B.Cardinetti,V.G.Benza,L.Molinari, Phys.Rev. B{\bf 52},
R5499(1995).
\bibitem{AW}A.G.Aronov,P.Wolfle,Phys.Rev. B {\bf 50},16574(1994).
\bibitem{MOW}E.W.Montroll,B.J.West, in:''Fluctuation Phenomena'',\\
E.W.Montroll,J.L.Lebowitz eds.,North Holland,Amsterdam,1987.
\bibitem{HA}H.Hiramoto,S.Abe,J.Phys.Soc.Jpn. {\bf57}.230(1988).
\bibitem{GKP}T.Geisel,R.Ketzmerick,G.Petschel, 
Phys.Rev.Lett. {\bf 66},1651(1991).
\bibitem{CD}J.T.Chalker,A.Dohmen,Phys.Rev.Lett. {\bf 75},4496(1995).
\bibitem{BF}L.Balents,M.P.A.Fisher,Phys.Rev.Lett. {\bf 76},2782(1996).
\bibitem{BFZ}L.Balents,M.P.A.Fisher,M.R.Zirnbauer,
Nucl.Phys. {\bf B483},601(1997).
\bibitem{MA}H.Mathur,Phys.Rev.Lett.{\bf 78},2429(1997).
\bibitem{BG}J.P.Bouchaud,A.Georges,Phys.Rep. {\bf 195},127(1990).
\bibitem{RJLS}R.del Rio,S.Jitomirskaya,Y.Last,B.Simon,
Phys.Rev.Lett. {\bf 75},117(1995).
\bibitem{MU} J.E.Muller, Phys.Rev.Lett. {\bf 68},385(1992).
\bibitem{CL}D.B.Chklovskii,P.A.Lee, Phys.Rev. B {\bf 46},18060(1993).
\end{thebibliography}
\end{document}